\documentclass{./emulateapj}
\newcommand{\mode}{emulateapj}
\usepackage{ifthen}
\usepackage{ifthen}
\newcommand{\mr}[1]{\mathrm{#1}}

\newcommand{\m}{$^{-1}$}

\ifthenelse{\isundefined{\nonaas}}{}{

\bibpunct{(}{)}{;}{a}{}{,}

  \renewenvironment{thebibliography}[1]{%
    \begin{oldthebibliography}{#1}%
      \setlength{\parskip}{0ex}%
      \parindent 0ex%
      \setlength{\itemsep}{0ex}%
  }%
  {%
    \end{oldthebibliography}%
  }

\def\apj{{ApJ}}			
\def\apjl{{ApJ}}		
\def\apjs{{ApJS}}		
\def\ao{{Appl.~Opt.}}		
\def\aap{{A\&A}}		
\def\mnras{{MNRAS}}		

}

\ifthenelse{\equal{\mode}{aastex}}{

\newcommand{\fighalf}{6in}
\newcommand{\dt}{deluxetable}
\newcommand{\trot}{\rotate}
}{
\submitted{Submitted February 12, 2007; Accepted June 18, 2007 for publication in \emph{The Astrophysical Journal}}

\newcommand{\fighalf}{3in}
\newcommand{\dt}{deluxetable*}
\newcommand{\trot}{}
}

\shortauthors{Mahdavi et al.}
\shorttitle{Abell 520}

\newcommand{\lfive}{2.13}
\newcommand{\ufive}{57 \pm 49}

\begin{document}

\title{A Dark Core in Abell 520$^\dagger$}
\altaffiltext{$\dagger$}{Based on observations obtained with
MegaPrime/MegaCam, a joint project of CFHT and CEA/DAPNIA, at the
Canada-France-Hawaii Telescope (CFHT) which is operated by the
National Research Council (NRC) of Canada, the Institute National des
Sciences de l'Univers of the Centre National de la Recherche
Scientifique of France, and the University of Hawaii.}

\author{Andisheh Mahdavi, Henk Hoekstra, Arif Babul, David D. Balam}
\affil{Department of Physics and Astronomy, University of Victoria,
Victoria, BC V8W 3P6, Canada}

\and

\author{Peter L. Capak}
\affil{California Institute of Technology, MC 105-24, 1200 E California 
Boulevard, Pasadena, CA 91125}

\begin{abstract}

The rich cluster Abell 520 ($z=0.201$) exhibits truly extreme and
puzzling multi-wavelength characteristics. It may best be described as
a ``cosmic train wreck.'' It is a major merger showing abundant
evidence for ram pressure stripping, with a clear offset in the gas
distribution compared to the galaxies (as in the bullet cluster 1E
0657-558). However, the most striking feature is a massive dark core
($721 h_{70} M_\odot/L_{\odot B}$) in our weak lensing mass
reconstruction.  The core coincides with the central X-ray emission
peak, but is largely devoid of galaxies. An unusually low mass to
light ratio region lies 500 kpc to the East, and coincides with a
shock feature visible in radio observations of the cluster. Although a
displacement between the X-ray gas and the galaxy/dark matter
distributions may be expected in a merger, a mass peak without
galaxies cannot be easily explained within the current collisionless
dark matter paradigm.  Interestingly, the integrated gas mass fraction
($\approx 0.15$), mass-to-light ratio ($220 h_{70} M_\odot/L_{\odot
B}$), and position on the X-ray luminosity-temperature and
mass-temperature relations are unremarkable. Thus gross properties and
scaling relations are not always useful indicators of the dynamical
state of clusters.

\end{abstract}


\section{Introduction}

Multi-wavelength studies of galaxy clusters have revealed a richly
textured intracluster environment, showing evidence of recent star
formation in cluster cores and powerfully disruptive events such as
AGN outbursts and mergers. Comparing observations of the hot
intracluster medium and the weak gravitational lensing of distant
sources has led to ever more precise constraints on the dark matter
distribution in these systems
\nocite{MiraldaEscude95,Allen02b,Dahle03,Smith05,Mahdavi07}(e.g. {Miralda-Escude} \& {Babul} 1995; {Allen} {et~al.} 2002; {Dahle} {et~al.} 2003b; {Smith} {et~al.} 2005; {Mahdavi} {et~al.} 2007).

Of particular interest is the recent discovery of offsets between the
dark matter distribution (as inferred from a weak lensing mass
reconstruction) and the hot gas distribution in X-ray emitting
clusters. The ``bullet cluster'' 1E 0657-558 demonstrates the power of
such a multi-wavelength approach: a 4500 km s\m\ cluster collision
results in stripping of the gas from the dark matter halo of the
smaller cluster, leaving a $\approx 100$ kpc projected separation
between the gas and lensing mass peaks
\nocite{Markevitch02,Clowe04}({Markevitch} {et~al.} 2002; {Clowe} {et~al.} 2004). This offset provides compelling
geometrical evidence for the existence of dark matter \nocite{Clowe06}({Clowe} {et~al.} 2006)
as well as limits on the dark matter self-interaction cross section
\nocite{Markevitch04}({Markevitch} {et~al.} 2004). Such mergers also offer an opportunity to study
gas physics through direct comparison of the shock properties with the
predictions of N-body and hydrodynamical simulations
\nocite{Hayashi06,Springel07}({Hayashi} \& {White} 2006; {Springel} \& {Farrar} 2007).  

These results suggest that studying {\it both} relaxed and merging
clusters can yield significant insights into the nature of dark matter
and the physics of the intracluster medium. With these goals in mind,
we have recently begun the Canadian Cluster Comparison Project
(CCCP)\footnote[1]{
\url{http://www.astro.uvic.ca/$\sim$hoekstra/CCCP.html}}, a
multi-wavelength survey of 50 massive clusters with gas temperatures
$> 5$ keV, of which roughly half are dynamically disturbed.

During our survey we discovered that Abell 520 ($z=0.201$; also known
as MS0451+02), a rich $10^{15} M_\odot$ merging system, exhibits truly
extreme multi-wavelength characteristics. In this paper we report on
the results of a weak lensing analysis and the subsequent comparison
with the optical and X-ray properties of the cluster. The data are
presented in \S\ref{sec:data}. In \S\ref{sec:analysis} we describe the
various mass constraints the data provide. In \S\ref{sec:discussion}
we discuss the implication for the cluster merger. The Appendix
provides a detailed comparison of the weak lensing results for the
various data sets that were used.  Throughout this Letter we assume
$H_0 = 70 h_{70}$ km s\m\ Mpc\m, $\Omega_0 = 0.3$, and
$\Omega_\Lambda=0.7$.

\section{Data Analysis}
\label{sec:data}

\subsection{Optical imaging}
\label{sec:optdata}

A major goal of the CCCP is the systematic study of the mass
distribution of clusters using weak gravitational lensing. To this end
we have obtained deep $g'$ and $r'$ wide field imaging data for a
sample of 30 clusters with the Canada-France-Hawaii Telescope (CFHT)
using MegaCam. The data for Abell 520 were obtained on November 13th,
2004. The observations consist of four 400s exposures in $g'$ and
eight 500s exposures in $r'$. For our weak lensing analysis we only
consider the $r'$ data because they are deeper and have better image
quality.

The $g'$ data are used to identify the cluster early type galaxies,
using their location on the well defined color-magnitude relation. We
select galaxies with $r<22$ and $g'-r'$ colors up to 0.25 magnitudes
bluer than the redward edge of the red sequence, which enables us to
map the cluster (red) light distribution and to compute the rest-frame
$B$-band luminosity. By removing these galaxies from our weak lensing
catalogs, we also reduce contamination by cluster members
\nocite{Hoekstra07}({Hoekstra} 2007).

The point spread function (PSF) introduces systematic changes in the
shapes of the galaxies used in the weak lensing analysis. Consequently
correcting for the effects of the PSF is a critical part of our
analysis. Current wide field imaging instruments such as Megacam
consist of a mosaic of chips and special care needs to be taken to
account for sudden jumps in the PSF properties when data from
different exposures are combined. To avoid such problems altogether,
the CCCP $r'$ data are obtained in two sets of exposures. Each set of
four 500s exposures is taken with small dithers. The two sets are
offset by approximately half a chip in each direction to fill in most
of the gaps between chips. The image quality of the data is excellent
with FWHMs of $0\farcs50$ and $0\farcs57$ as measured from the stacked
sets. Each set is analyzed separately, which ensures that all data
originate from the same chip.  The resulting catalogs with shape
parameters are then combined (see the Appendix for details).

Abell~520 has also been observed using Subaru using the Suprime-Cam in
the $i'$ and Cousins $R_C$ bands. These data were used by
\nocite{Okabe07}{Okabe} \& {Umetsu} (2007) in their weak lensing analysis of the cluster.
In order to compare to our CFHT observations, we retrieved these data
from the Subaru archive and analyzed the data ourselves. The six $R_C$
images were obtained on October 15 and 19 2001 and have integration
times of 300s, with a seeing range from $0\farcs52$ to $0\farcs65$.
The seven $i'$ images were obtained on October 19th 2001 and November
17th 2001 and have integration times of 240s. The image quality of
these data is excellent as well, with the seeing ranging from
$0\farcs43$ to $0\farcs57$.

The Subaru data were taken with relatively large offsets from exposure
to exposure. Furthermore, the PSF showed large variations across the
field of view and from image to image. We found that by stacking the
data, the resulting PSF pattern could not be corrected for to the
level required for our analysis. Instead, we chose to analyze each
exposure separately, and to combine the weak lensing shape catalogs
instead. This approach significantly reduced residual systematics (to
a negligible level).

\subsection{Weak lensing shear measurement}

Our weak lensing analysis follows the procedures outlined in
\nocite{Hoekstra07}{Hoekstra} (2007) and is based on the algorithm described in
\nocite{Kaiser95}{Kaiser} {et~al.} (1995) with modifications described in \nocite{Hoekstra98}{Hoekstra} {et~al.} (1998), and
\nocite{Hoekstra00}{Hoekstra} {et~al.} (2000). We refer the interested reader to those papers. We
note that this implementation has been tested rigorously
\nocite{Hoekstra98,Heymans06}(e.g. {Hoekstra} {et~al.} 1998; {Heymans} {et~al.} 2006) and has been shown to be accurate
at the few percent level. The first step in the analysis is the
identification of the objects, which are subsequently analyzed. Our
pipeline results in a catalog of ellipticities for the faint galaxies
we use in our analysis. These shapes have been corrected for PSF
anisotropy and the size of the PSF.

\begin{figure}
\begin{center}
\begin{tabular}{cc}
\resizebox{!}{\fighalf}{\includegraphics{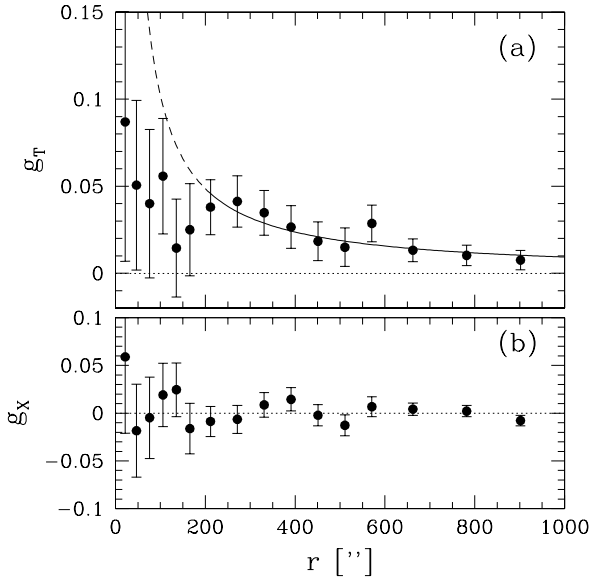}}&
\end{tabular}
\figcaption{(a) Average tangential distortion as a function of
distance from the peak of the X-ray emission using measurements from
our Megacam $r'$ data (20<$r'$<25). The line is the best fit singular
isothermal sphere model, fitted to data at radii larger than 200'';
the corresponding velocity dispersion is $1028 \pm 80$ km s\m. (b) the
signal when the phase of the distortion is increased by $\pi/2$. If
the signal observed in the upper panel is due to gravitational
lensing, $g_x$ should vanish, as is observed.
\label{fig:gtprof}}
\end{center}
\end{figure}

Figure~\ref{fig:gtprof} shows the resulting tangential distortion as a
function of distance from the peak of the X-ray emission. To measure
the signal we used the measurements obtained from the Megacam $r'$
data, selecting galaxies with $20<r'<25$. We detect a significant
lensing signal, but as discussed below, the mass distribution in the
central region is complicated, resulting in a decrease in the
tangential distortion. 
To relate the observed lensing signal to an estimate of the mass
requires an estimate of the mean source redshift distribution. To do
so, we use the photometric redshift distribution from \nocite{Ilbert06}{Ilbert} {et~al.} (2006),
which are based on the CFHT Legacy Survey Deep fields.  The strength
of the lensing signal can be characterized by $\beta=\max[0,D_{ls}
/D_s]$, where $D_{ls}$ and $D_s$ are the angular diameter distances
between the lens and the source, and the observer and the source. We
note that the value for $\beta$ varies for the data sets considered
here. In the Appendix we present a detailed comparison of the various
catalogs, including an estimate for $\beta$ for each. The selection of
CFHT-detected objects used to compute the signal presented in
Figure~\ref{fig:gtprof} results in $\beta=0.60$. For reference, we fit
a singular isothermal sphere model to the measurements at radii larger
than 200'' (solid line in Figure~\ref{fig:gtprof}). The best-fit
velocity dispersion is $\sigma=1028 \pm 80$ km s\m.

For the surface density reconstruction use the direct inversion
algorithm from \nocite{Kaiser93}{Kaiser} \& {Squires} (1993), which works well for the wide field
imaging data used here. The results are presented in
Figure~\ref{fig:bigpic}. We used the object catalog with all detected
galaxies ($\beta = 0.59$) for this reconstruction and for all masses
reported in the main section of the paper. The shape parameters of
objects that were detected in multiple filters were averaged as
discussed in the Appendix. Figure~\ref{fig:bigpic} also shows the
optical, X-ray, and lensing maps for Abell 520.  The resolution of the
mass reconstruction is limited by the number density of source
galaxies in these ground based observations, and the FWHM of the
Gaussian smoothing kernel is $\sim 60''$.

\begin{\dt}{lrrrrrrrrr}
\tablewidth{0in}
\trot
\tablecaption{X-ray properties of the lensing peaks \label{tbl:peaks}}
\tablehead{\colhead{Peak} & \colhead{RA(J2000)} & \colhead{DEC(J2000)}
& \colhead{$k T_X$} & \colhead{$Z_X$} & \colhead{z} & \colhead{$N_H$} & 
\colhead{Norm} & \colhead{$L_X$} & \colhead{$\ell$} \\
& & & \colhead{keV} & \colhead{$Z_\odot$} & & \colhead{$10^{22}$ cm$^{-2}$} & \colhead{$10^{-18}$ cm$^{-5}$} & \colhead{10$^{43}$ erg s\m} & \colhead{Mpc} }
\startdata
1 & 04:54:19.94 & +02:57:45.4 & 8.0$_{-1.6}^{+2.4}$ & $<2$                  &$=0.201$        & $=0.018$                 & $0.9_{-0.3}^{+0.2}$  & 0.5  & 1.1 \\ 
2 & 04:54:14.08 & +02:57:08.9 &13.5$_{-2.5}^{+3.4}$ & $<0.31$               &$=0.201$        & $=0.018$                 & $3.8_{-0.2}^{+0.1}$  & 1.8  & 1.6 \\
3 & 04:54:10.41 & +02:55:20.4 & 9.8$_{-0.6}^{+0.7}$ & $0.40_{-0.10}^{+0.10}$&$=0.201$        & $=0.018$                 & $11.0_{-0.2}^{+0.2}$ & 5.5  & 2.0 \\
4 & 04:54:03.81 & +02:53:30.3 & 6.3$_{-0.6}^{+0.8}$ & $0.38_{-0.22}^{+0.23}$&$=0.201$        & $=0.018$                 & $3.0_{-0.2}^{+0.1}$  & 1.5  & 1.7 \\
5 & 04:54:20.05 & +02:55:31.5 &10.7$_{-2.1}^{+2.8}$ & $<2$                  &$=0.201$        & $=0.018$                 & $1.9_{-0.2}^{+0.1}$  & 1.0  & 1.7 \\
Cluster& \nodata& \nodata     & 9.3$_{-0.5}^{+0.4}$ & $0.36_{-0.05}^{+0.05}$&$0.209_{-0.01}^{+0.01}$ & $0.018_{-0.005}^{+0.005}$& $75.5_{-0.8}^{+0.8}$& 37.8 & 2.0 \\
\enddata
\tablecomments{We list the position of each lensing peak, along with
the best-fit MEKAL X-ray temperature and metallicity fit to data
extracted within a 150 kpc radius circular aperture. ``Cluster'' is a
circle centered on peak 3 with a 710 kpc radius. The redshift was
fixed at the optical value, and the absorbing column was set to the
best-fit value for the entire cluster. The plasma model normalization
is $\int n_e n_H dV / 4 \pi D^2$, where $n_e$ and $n_H$ are the
electron and proton space densities, and $D$ is the comoving distance
to the cluster. The X-ray luminosity is measured in the 0.5-2.0 keV
band.  The effective column along the line of sight for each peak,
$\ell$ is conservatively estimated (see \S\ref{sec:data}). Errors are
$1 \sigma$; upper limits are at 90\% confidence.}
\end{\dt}

\subsection{Optical spectroscopy}
\label{sec:redshift}

Abell 520 was part of the Canadian Network for Observational Cosmology
(CNOC) survey \nocite{Yee96,Carlberg96}({Yee} {et~al.} 1996; {Carlberg} {et~al.} 1996). \nocite{Proust00}{Proust} {et~al.} (2000) independently
measured redshifts for 29 galaxies within 1 Mpc of the X-ray center of
Abell~520. We create a composite catalog by merging both redshift
surveys; the detailed properties of this catalog will appear in a
subsequent paper. 

After a 3-$\sigma$ clipping of the redshifts, we find a well defined
velocity peak with 71 member galaxies within 1 Mpc. The mean velocity
is $c z = 60307 \pm 155$ km s\m, or $z=0.201$.  The velocity
dispersion within the same radius, corrected for $(1+z)$ cosmological
broadening, is $1120 \pm 75$ km s\m. This value is in excellent
agreement with the weak lensing value of $1028\pm 80$ km s\m, based a
singular isothermal sphere model fit to the signal.

\subsection{X-ray data}
\label{sec:xrayobs}

We reanalyze ObsID 4215, the 67ks Chandra X-ray observation originally
described by \nocite{Markevitch05}{Markevitch} {et~al.} (2005), using the CIAO 3.3 reduction
software and CALDB version 3.2.0 along with the standard background
analysis tools.

We reapply the standard reduction pipeline, including the charge
transfer inefficiency correction \nocite{Townsley00}({Townsley} {et~al.} 2000), and calculate
properly weighted response matrices using the CIAO \verb-mkwarf- and
\verb-mkacisrmf- tasks. Point sources in the field are detected by the
CIAO \verb-wavdetect- package and masked. We use the CIAO blank sky
fields to remove the particle background, and then subtract a
$2\arcmin \times 2\arcmin$ area free of cluster emission to account
for the cosmic and soft X-ray backgrounds. The X-ray center of the
diffuse emission was taken to be the centroid of the cluster emission.

To analyze the spectra, we fit absorbed single-temperature MEKAL
models. The X-ray measurements, including temperatures and
luminosities, are shown in Table \ref{tbl:peaks}, and are consistent
with the values reported by \nocite{Markevitch05}{Markevitch} {et~al.} (2005) for the cluster.

Because of the complex, irregular mass distribution, spherical
deprojection of the cluster to arrive at the gas mass is
inappropriate. Instead, using the X-ray luminosity of the gas, we
derive an upper limit on the gas mass \emph{integrated within an
aperture and along the line of sight}. Consider the Cauchy-Schwartz
inequality
\begin{equation}
\left( \int \rho_g dV \right)^2 \le \int \rho_g^2 dV \int dV
\end{equation} 
where $\rho_g$ is the gas density and $dV$ is the volume element along
the line of sight.  The left hand side is the square of the gas mass,
while the right hand side is proportional to the X-ray luminosity
times the emitting volume. To place an upper limit on the gas mass,
one only requires an estimate of the column along the line of sight,
$\ell$, which we take to be the value appropriate for a 1 Mpc radius
sphere, $2 (1-d^2)^{1/2}$, where $d$ is the projected distance of the
extracted region from the X-ray center. This results in a generous
upper limit because the visible extent of X-ray emission is only
$\approx 600$ kpc. Departures from sphericity can be accounted for in
a straightforward manner: stretching the emitting column $\ell$ by a
factor $\alpha$ gives a $\sqrt{\alpha}$ increase in the upper
limit. We neglect departures from isothermality\footnote{
The temperature dependence of the gas mass is negligibly affected by
the free-free Gaunt factor $g_{ff}$ \nocite{Rybicki86}({Rybicki} \& {Lightman} 1986).}
: at the high temperatures reported here
the luminosity varies only as $T^{1/2}$, and the gas mass upper limit
would vary only as a $T^{1/4}$.

\section{Results}
\label{sec:analysis}

\subsection{Morphology}

\begin{figure*}
\begin{center}
\begin{tabular}{cc}
\multicolumn{2}{c}{
\resizebox{5in}{!}{\includegraphics{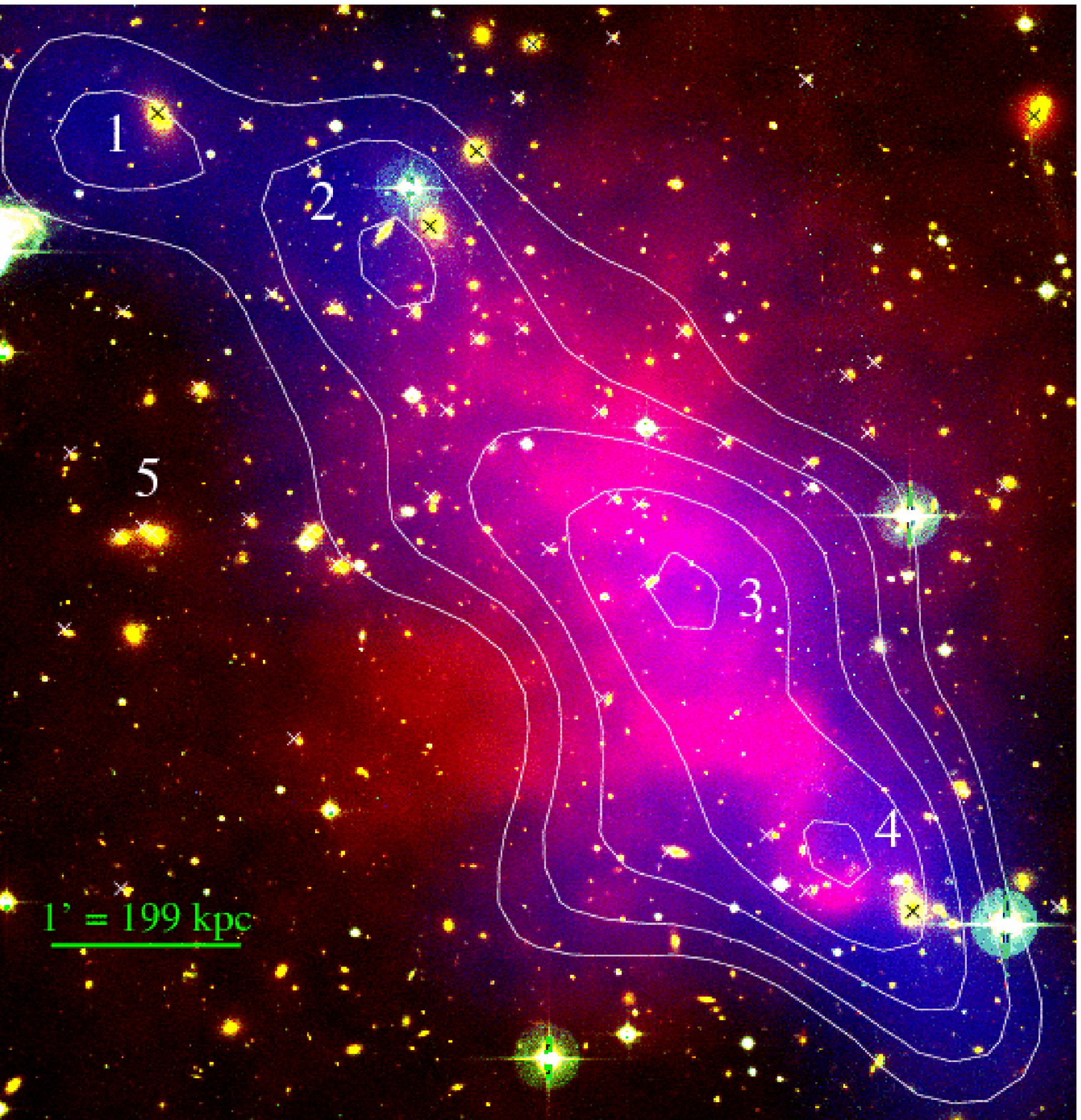}}} \\
\resizebox{2.5in}{!}{\includegraphics{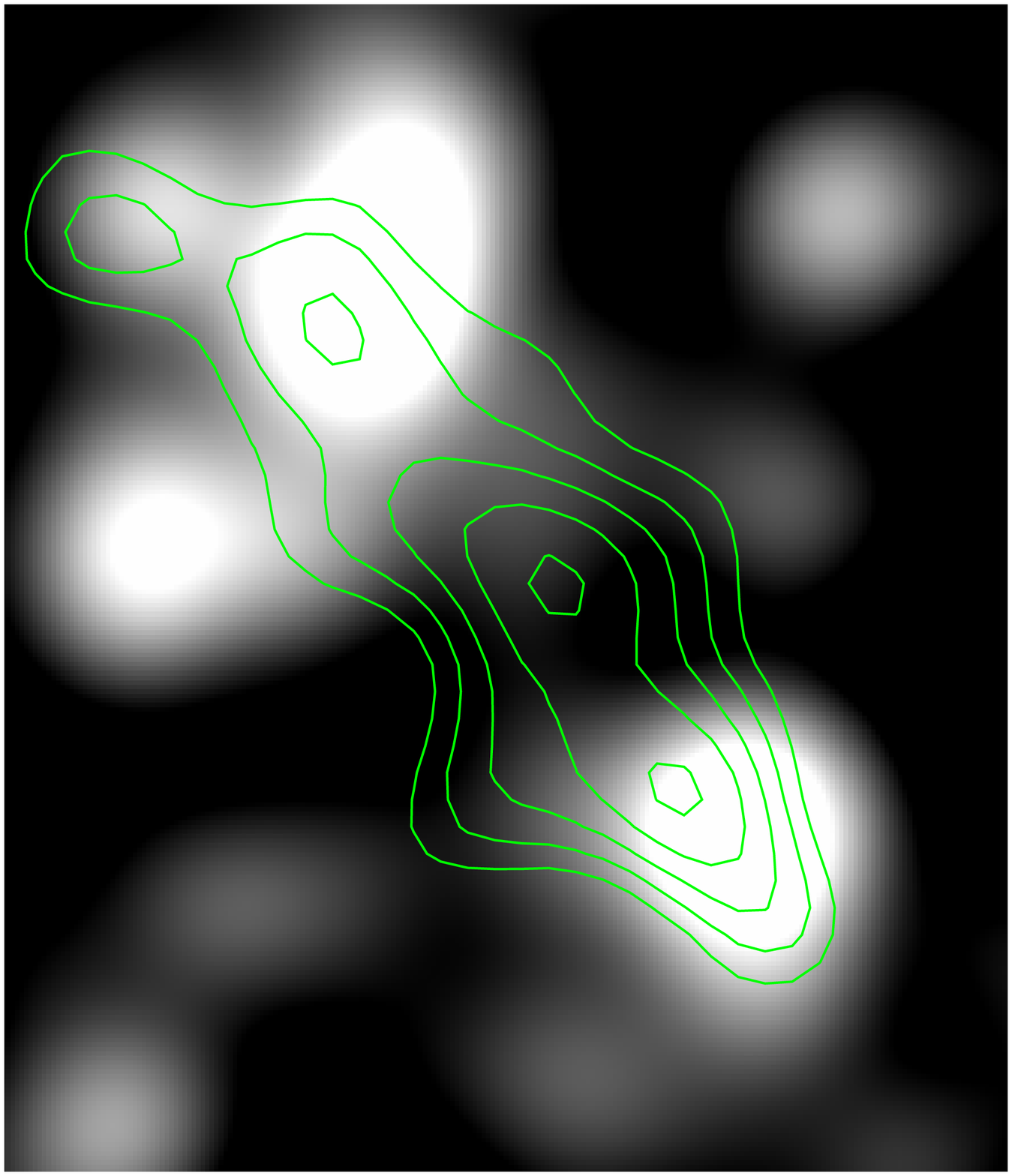}} &
\resizebox{2.5in}{!}{\includegraphics{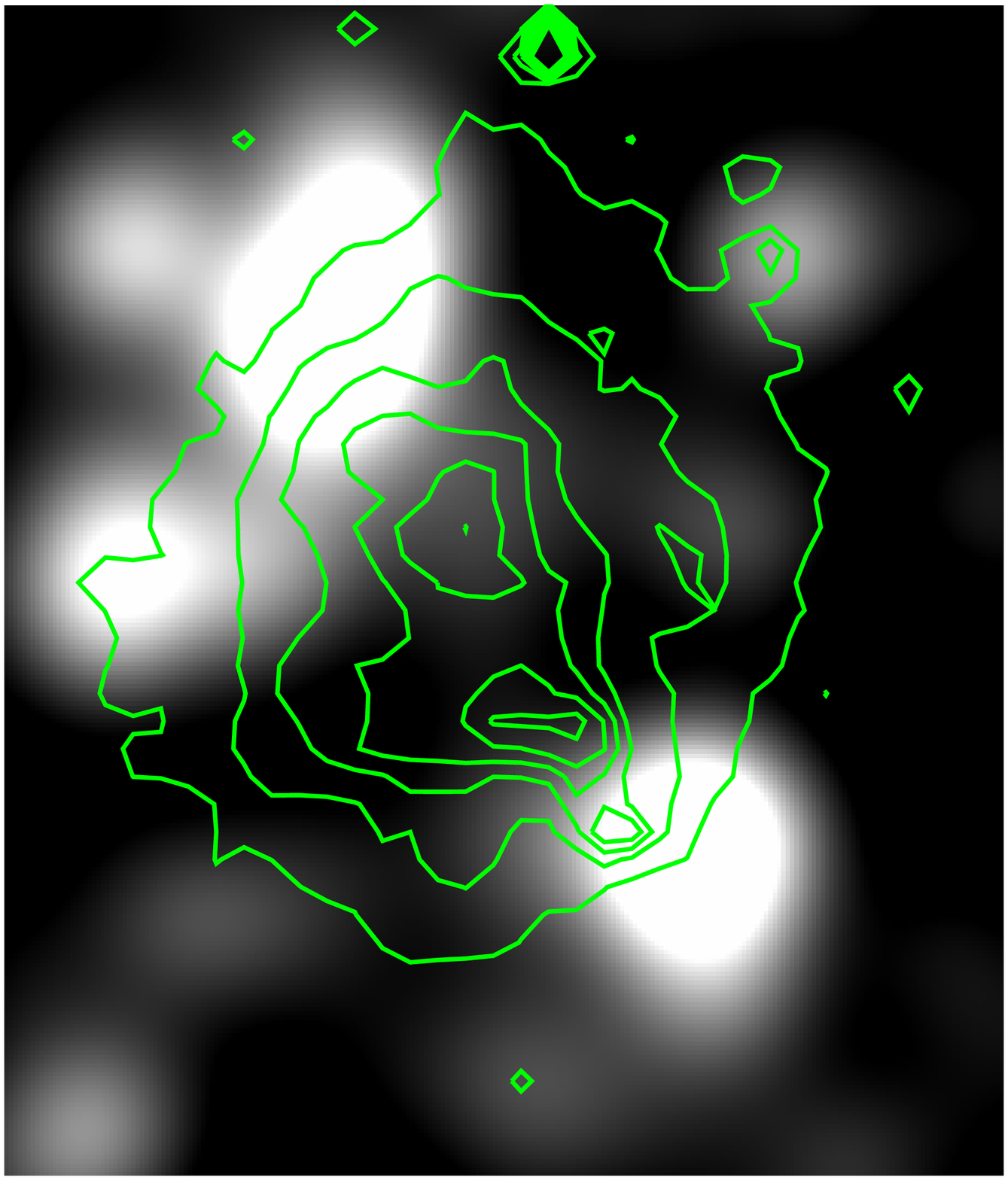}} 
\end{tabular}
\figcaption{(a) Central $6.4\arcmin \times 6.4\arcmin$ of Abell
520, showing the CFHT image, the diffuse Chandra X-ray emission (red),
and the lensing surface mass density (blue + 3, 3.5, 4, 4.5, and
5$\sigma$ contours determined from a bootstrap
analysis). Spectroscopically confirmed member galaxies are marked with
an X; red-sequence galaxies appear orange. (b) Red light distribution
together with lensing contours from (a). (c) Same as (b), but with
X-ray contours. Note the absence of galaxies in the central lensing
peak.
\label{fig:bigpic}}
\end{center}
\end{figure*}

Weak gravitational lensing studies of clusters allow us to reconstruct
an `image' of the matter distribution that is independent of
assumptions regarding the geometry or dynamical state of the cluster.
This is a unique feature and crucial when studying complex systems
such as Abell~520, as demonstrated here.

The striking incongruities in the stellar, gas, and dark matter
distribution make this cluster a truly unique example of a massive,
perhaps even multiple merger. The lensing map indicates that the
cluster is aligned in the NE-SW direction, with four 5$\sigma$ peaks
which we number 1-4. The NE-SW direction is corroborated by the
Chandra data, which show an ``arm'' extending SW from the main X-ray
emitting region towards peak (4). The X-ray arm and accompanying shock
were previously discussed by \nocite{Markevitch05}{Markevitch} {et~al.} (2005).  We also identify a
fifth region (``peak 5''), with high red luminosity but low lensing
mass. The properties of the five regions appear in Tables
\ref{tbl:peaks} and \ref{tbl:masses}. The masses were determined
directly from the lensing signal using a one-dimensional mass
reconstruction \nocite{Hoekstra07}(a.k.a. aperture mass densitometry; for details
see {Hoekstra} 2007).

Peaks 3 and 5 are the most remarkable features of Abell 520 and
deserve extended attention. The significance of peak 3 is comparable
to the other mass peaks, yet its mass-to-light ratio is anomalously
high, $721 h_{70} M_\odot / L_{B\odot}$, compared to $\sim 200$ for
typical groups and clusters \nocite{Girardi02}({Girardi} {et~al.} 2002).  Conversely, peak 5 has
a mass-to-light ratio that is strikingly low, $\ufive h_{70} M_\odot /
L_{\odot B}$. Furthermore peak 5 also coincides with diffuse radio
emission, as is readily apparent through a comparison of Figure
\ref{fig:bigpic}a with Figure 1 of \nocite{Markevitch05}{Markevitch} {et~al.} (2005). We discuss
possible explanations for this remarkable configuration below
(\S\ref{sec:discussion}).

None of the three brightest cluster galaxies (BCGs; $M_B < -22$) are
within peak 3; rather they are 690, 490, and 520 kpc from the center,
located in peaks 1, 2, and 4, respectively. The brightest galaxy
within 150 kpc of peak 3 has $M_B=-20.3$, hardly brighter than $L^*$
at $z=0.2$ \nocite{Croton05,Dahlen05}({Croton} {et~al.} 2005; {Dahlen} {et~al.} 2005). We know of no other groups or
clusters without a $M_B < -20.3$ member within 300 kpc of coincident
X-ray \emph{and} lensing peaks.

\subsection{Reality of the Dark Peak}

Here we argue that peak 3 contains mostly dark matter and is
physically associated with the cluster. First, the high $721 h_{70}
M_\odot / L_{\odot B}$ mass-to-light ratio argues for a deficit of
stars in the peak. Second, the upper limit on the gas mass fraction,
$f_g<0.17$, suggests that $\gtrsim 85\%$ or $2.8 \times 10^{13}$
$M_\odot$ of material has to be dark. If e.g. half the material within
peak 3 were hot gas stripped from the plentiful reservoirs of the
merger precursors, the observed X-ray luminosity would be $\approx 50$
times greater than the measured $5.5 \times 10^{43}$ erg s\m.

Chance superpositions are a factor to be considered; a superposed
background structure could mimic a dark peak. However, the
multi-wavelength data firmly rule out this hypothesis for a number of
reasons. First, the X-ray spectrum of the gas exhibits the rest-frame
6.6 keV Fe line complex, which constrains the redshift of the X-ray
emission within 1\arcmin\ of peak 3 to be $z = 0.21 \pm 0.01$.
Second, the brightest three galaxies in peak 3 all fall in the cluster
red sequence, and are spectroscopically verified cluster members
according to the CNOC data. Finally, a background cluster along the
line-of-sight would result in a detectable excess of galaxies at the
position of the peak, which is not observed. The latter argument might
be countered by considering a very high redshift $(z>1)$
cluster. However, the amplitude of the lensing signal, and its
dependence with limiting $r'$ magnitude for the source galaxies,
clearly argue against such a high redshift.

A final source of concern is that peak 3 could arise from the
superposition of two separate dark halos centered on peaks 2 and 4. If
the two halos have recently passed through each other, the material in
the overlapping portion could well give rise to a surface density
enhancement. We estimate the amount of enhancement by approximating
the matter distribution in peaks 2 and 4 as \nocite{NFW}{Navarro} {et~al.} (1997, NFW) spheres
with density profiles \mbox{$\rho_0 (r/$100~kpc$)^{-1}
(1+r/$100~kpc$)^{-2}$}, truncated at a distance of 2 Mpc from the
X-ray center, with $\rho_0$ determined from the lensing data
itself. We find that the total amount of mass within 150 kpc of peak 3
purely from the two intersecting NFW spheres is $\approx 2 \times
10^{12} M_\odot$. This is an order of magnitude smaller than the dark
mass implied by the X-ray and lensing observations. This result is
insensitive to even large changes in the NFW concentration.

The available evidence therefore suggests that peak~3 is a distinct
physical structure associated with Abell~520.

\subsection{Previous studies}

\nocite{Dahle02}{Dahle} {et~al.} (2002) include Abell 520 in their weak lensing study of
38 X-ray emitting clusters. They use the University of Hawaii
88\arcsec\ telescope with the 8K mosaic camera; the data have
$0\farcs9$ seeing, compared to $0\farcs5$ for the Subaru and CFHT
data.  While the \nocite{Dahle02}{Dahle} {et~al.} (2002) map for Abell 520 does not recover all
the features we see in our mass reconstruction, its most significant
peak is centered very near (within 50 kpc) of our peak 3.

We note that our result differs from the analysis presented by
\nocite{Okabe07}{Okabe} \& {Umetsu} (2007), also based on the Subaru data used here. They detect a
significant mass at the position of peak 5, whereas peak~3 is not
particularly pronounced. We believe that a problem with the PSF
anisotropy correction is the source of the difference.  \nocite{Okabe07}{Okabe} \& {Umetsu} (2007)
analyzed the stacked data, which is problematic for the Abell 520
Subaru observations (see \S\ref{sec:optdata}).

To examine the robustness of the final surface density map, we also
measured the weak gravitational lensing signal from the $R$, $r'$, and
$i'$ data separately. We find that our results are robust, most
notably the peak in the mass distribution that coincides with the peak
of the X-ray emission (see the Appendix for more details regarding the
comparison of the weak lensing measurements).

\begin{\dt}{lccccccccc}
\tablewidth{0in}
\tablecaption{Masses and Mass-to-light Ratios \label{tbl:masses}}
\tablehead{\colhead{Peak} & \colhead{$M_\mr{wl}$} & \colhead{$L_B$}
& $\Upsilon$ & $M_g$ & $f_g$ \\
 & \colhead{$h_{70}^{-1} 10^{13} M_\odot$} &  \colhead{$ h_{70}^{-2} 10^{11} L_{\odot B}$}
& $h_{70} M_\odot / L_{\odot B}$ & $h_{70}^{-2.5} 10^{13} M_\odot$ & $h_{70}^{-1.5}$}
\startdata
1      & $3.73 \pm 0.99$ & $1.59$ & $234 \pm 62$ & $<0.12$ & $<0.05$ \\
2      & $3.60 \pm 1.04$ & $4.22$ & $85  \pm 25$ & $<0.28$ & $<0.12$ \\
3      & $4.40 \pm 1.09$ & $0.61$ & $721 \pm 179$& $<0.52$ & $<0.17$ \\
4      & $4.82 \pm 0.89$ & $3.56$ & $135 \pm 25$ & $<0.25$ & $<0.07$ \\
5      & $1.22 \pm 1.06$ & \lfive &  $\ufive$    & $<0.22$ & $<1$    \\
Cluster& $49.98\pm 5.47$ &$21.57$ & $232 \pm 25$ & $<6.44$ & $<0.15$ \\ 
\enddata
\tablecomments{Shown are the projected lensing mass, the blue
rest-frame luminosity and the mass-to-light ratio for each lensing
peak, measured within a 150 kpc aperture from the combined $r'$, $i'$,
and $R$ catalogs. ``Cluster'' is a circle centered on peak 3 with a
710 kpc radius.  $M_g$, the cylindrical gas mass integrated along the
line of sight within the same aperture, and $f_g$, the corresponding
baryon fraction, are model-independent (see \S\ref{sec:data}). Errors
are 1$\sigma$; upper limits are at 90\% confidence.}
\end{\dt}

\section{Interpretation}
\label{sec:discussion}

\subsection{Merger characteristics}

The data suggest a head-on merger of roughly equal mass clusters along
the NE-SW axis, with a possible secondary E-W merger related to peaks
3 and 5.  Our interpretation is supported by the significant velocity
structure associated with each of the mass peaks. The rest-frame
line-of-sight velocities of the brightest galaxies in peaks 1, 2, 3, 4,
and 5 are 67 km s\m, -600 km s\m\, +700 km s\m, +400 km s\m, and -1300
km s\m, respectively. The peaks are therefore clearly offset from each
other in velocity space. Second, as is the case for the famous
``bullet cluster'' 1E 0657-558 \nocite{Markevitch02,Clowe06}({Markevitch} {et~al.} 2002; {Clowe} {et~al.} 2006), we find
that the X-ray emission is offset from the galaxy distribution.  This
offset is expected for ram pressure stripping, and is seen in gas+dark
matter simulations of merging clusters \nocite{Poole06}({Poole} {et~al.} 2006). Furthermore,
we observe low gas fractions in peaks 1 and 4 (Table
\ref{tbl:masses}).  The post-shock Mach number derived by
\nocite{Markevitch05}{Markevitch} {et~al.} (2005) for the peak 4 X-ray emission is $\approx 2.2$,
corresponding to a velocity of $\approx 1000$ km s\m. Combined with
the peak 4 line-of-sight velocity, this suggests an inclination of
$\approx 60\degr$ for the cluster, and leads to a merging timescale of
$\approx 1$ Gyr.

Despite its extreme characteristics, Abell 520 would appear typical if
its global integrated properties were considered alone (Tables
\ref{tbl:peaks} and \ref{tbl:masses}). The integrated gas fraction,
$\lesssim 0.15$, is normal for clusters of this mass, as is its
mass-to-light ratio, $220 M_\odot/L_{\odot B}$
\nocite{Girardi02}({Girardi} {et~al.} 2002). 

The peak 3 mass-to-light ratio is larger than some previously
published ``extreme'' values once they are corrected to our
$\Lambda$CDM cosmology. For example, \nocite{Gray02}{Gray} {et~al.} (2002) claim a $529
M_\odot/L_{\odot B}$ for Abell 901b.  What sets Abell 520 apart from
previous work is the distinct lack of a BCG at the center of the dark
peak. In the case of clusters such as Abell 901b, the dark matter peak
and the BCG are well aligned; but in Abell 520, all the bright
galaxies are $>400$ kpc removed from the center of the dark
peak. Abell 520 is unique because the lensing signal and the X-ray
emission coincide in a region that lacks bright galaxies.

Remarkably, the cluster falls on the X-ray luminosity-temperature
relation \nocite{Reiprich02,Lumb04}({Reiprich} \& {B{\" o}hringer} 2002; {Lumb} {et~al.} 2004) and mass-temperature relation
\nocite{Vikhlinin06}({Vikhlinin} {et~al.} 2006). Thus while departures from scaling relations may
be useful for gauging deviations from equilibrium, clusters consistent
with these relations are not necessarily close to equilibrium
or``preferred'' in any other sense.

\subsection{Separating Dark from Light}

What truly distinguishes Abell 520 from other extreme clusters is the
presence of a central dark region (peak 3) with almost no galaxies,
and of a corresponding ``light'' region (peak 5) largely \emph{devoid
of dark matter}, i.e. consistent with being almost entirely baryonic
based on both the X-ray and the optical data. Unlike the gas, both
cold dark matter (CDM) and galaxies ought to be collisionless, and
therefore even violent events should not be able to separate these two
components.
Having rejected the possibility of a background cluster, we now
consider other possible explanations.

If we were to add all mass components of both peaks 3 and 5, we would
have a peak with an ordinary mass-to-light ratio of $205 h_{70}
M_\odot/L_{B\odot}$ and baryon fraction $<0.17$. Given that the
velocities of the galaxies in peak 5 are systematically blueshifted
with respect to the cluster mean, it is highly likely that peak 5 is a
dynamically distinct subsystem.  Furthermore, peak 5 is coincident
with both very hot X-ray emitting gas \emph{and} diffuse radio
emission, which together make a very strong argument for the existence
of a secondary shock in the cluster.  For these reasons we consider
the possibility that peak 3 and 5 shared a common merger precursor,
which passed through the cluster from the west to the east in
projection on the sky.

The scenario most consistent with the CDM paradigm is that peak 3
occurred as a result of complex evolution during a multiple
merger. Recent gas and dark matter simulations show that during such
mergers, distortion and elongation of the original dark matter halos
is commonplace \nocite{Poole06}({Poole} {et~al.} 2006).  Recently, \nocite{Sales07}{Sales} {et~al.} (2007) showed that
under the right conditions, collapsed satellites falling in at late
times could be ejected via three body interactions after passing
through the center of the host halo. Perhaps a similar process
succeeded in ejecting the galaxies in peak 5 from their host halo,
peak 3. It is unclear, however, whether the \nocite{Sales07}{Sales} {et~al.} (2007) process
could operate on a cluster scale, with galaxies embedded in a massive
halo. We are presently conducting simulations to test this hypothesis.

A more intriguing possibility is that the dark matter concentration at
the center-of-mass of the merger is due to the collisional deposition
of dark matter.  If the dark and light components (peaks 3 and 5) had
a common origin, and the dark matter was collisionally stripped from
this precursor, then Abell 520 is a counterexample to the bullet
cluster. \nocite{Markevitch04}{Markevitch} {et~al.} (2004) used the coincidence of the galaxy and
lensing peaks in the bullet cluster to set an \emph{upper} limit on
the dark matter interaction cross section. For Abell 520, we use their
equations 18-19 to estimate the necessary dark matter interaction
cross section to produce peak 3.

Suppose that each of peaks 1, 2, 4, and 5 each contributed $\sim 25\%$
of the total mass observed in the central dark peak. The chief unknown
in the calculations is the surface density of each subcluster as
viewed by an oncoming particle. For Abell 520, using the mass for peak
3 and assuming an effective depth along the merging axis of 150 kpc,
we estimate $\Sigma_m \approx 0.066 \pm 0.016$ g/cm$^2$, yielding a
cross section of $\sigma_\mr{dm}/m_\mr{dm} \approx 0.25/0.066\approx
3.8 \pm 1.1 $ cm$^2$ g\m, well above the upper limit of 1 cm$^2$ g\m\
derived for the bullet cluster. The depth could hardly be smaller than
150kpc, and greater depths would only lead to higher cross
sections. Other constraints from cluster mass profiles suggest
$\sigma_\mr{dm} < 0.1$ cm$^{2}$ g\m
\nocite{Meneghetti01,Dahle03b,Arabadjis05}({Meneghetti} {et~al.} 2001; {Dahle} {et~al.} 2003a; {Arabadjis} \& {Bautz} 2005); these do not take the
central baryon distribution into account. We note that light dark
matter candidates such as axions and supersymmetric weakly interacting
massive particles (WIMPs) have self-interaction cross sections many
orders of magnitude below all the values discussed here
\nocite{Kamionkowski02}({Kamionkowski} 2002).

The $\sigma_\mr{dm}$ value is an order-of-magnitude estimate, but any
detailed corrections to the estimate must also be reflected in the
upper limit derived for the bullet cluster. We caution that the
measurement is sensitive to the surface density of the cluster along
the merging direction, something that is uncertain in our current maps
but will improve with planned higher resolution Hubble Space Telescope
observations. Differences in the merger impact parameter could be
invoked to explain why Abell 520 and the bullet cluster yield
different constraints on the cross-section; such a discussion is
beyond the scope of this paper.

A final possibility is that we are observing a filament unrelated to
the merger and by chance elongated almost exactly along the line of
sight. The filament must be sparse enough that it fails to produce any
significant galaxy concentration, sparse enough that it has little
detectable X-ray emission, and long enough to produce the observed
$4.4 \times 10^{13} M_\odot$ mass concentration. This hypothesis can
be tested using our upcoming Sunyaev-Zel'dovich observations, which
together with the X-ray data ought to reveal the structure of the gas
along the line of sight.

\section{Conclusions}

The merging cluster Abell 520 demonstrates the power of
multi-wavelength techniques in revealing extreme phenomena in clusters
of galaxies. Our study highlights the usefulness of studying unusual
clusters. The most remarkable finding of this work is the evidence for
a massive, 721 $ h_{70} M_\odot / L_{\odot B}$ dark core that
coincides with the peak in the X-ray emission, but is surprisingly
devoid of bright galaxies.  A ``luminous'' region containing little or
no dark matter lies 500 kpc to the east.

To test the robustness of the weak lensing analysis, we perform the
first detailed study of multi-telescope, multi-bandpass weak lensing
observations of clusters of galaxies. As shown in the Appendix, our
joint analysis of the Subaru and CFHT data demonstrates that the dark
peak in Abell 520 is not the result of instrumental effects, and that
our methodology yields consistent results in the $R$, $r'$, and $i'$
bandpasses.

Using a model-independent analysis of the X-ray and optical data, we
argue that the dark peak is physically associated with the cluster,
and that it consist of $\gtrsim 85\%$ dark matter. We estimate a
timescale of 1 Gyr for the merger; together with the shock velocity
derived by \nocite{Markevitch05}{Markevitch} {et~al.} (2005), we derive an inclination of $\approx
60\degr$ for the cluster. 

Abell 520 lies on the cluster mass-temperature and
luminosity-temperature relations. Therefore consistency with cluster
scaling relations is not necessarily an indicator that a cluster of
galaxies is relaxed.

We consider possible mechanisms for separating the dark matter from
the galaxies. Two possibilities stand out: (a) the galaxies originally
in the dark core could have been ejected through a multiple-body
interaction within the merging system; or (b) allowing for weakly
self-interacting dark matter, the dark peak was deposited as a result
of dark matter collisions during the merger impact; the required
self-interaction cross-section would be $3.8 \pm 1.1$ cm$^2$ g\m.
N-body simulations and higher resolution optical, X-ray, and radio
data are required to distinguish among these and other possible
explanations for this ``cosmic trainwreck.''

\acknowledgments

We thank the referee for comments that significantly improved the
paper. We are especially grateful to Howard Yee and Erica Ellingson
for sharing the Abell 520 CNOC velocity data. We thank Maxim
Markevitch, Mike Gladders, Neal Katz, Christoph Pfrommer, Pat Henry,
Julianne Dalcanton, Kathleen Mahdavi, Gil Holder, and Greg Poole for
insightful discussions. AB and HH acknowledge support from NSERC. AB
also acknowledges support from the Leverhulme trust in the form of a
visiting professorship at Oxford and Durham Universities. HH also
acknowledges support from CFI, BCKDF and CIfAR. Additional research
funding was provided by J. Criswick.

\appendix
\section{Detailed Comparison of CFHT and Subaru Multicolor Data}
The combination of Megacam $r'$ data and Subaru $R_C$ and $i'$ data
provides a unique opportunity to study the weak lensing signal as a
function of telescope and bandpass filter. Such a comparison is useful
because it provides a test of our ability to remove observational
distortions, which may be telescope or filter dependent. Below we
demonstrate that our reduction procedure yields consistent mass
reconstructions for all the instrument/filter combinations. In the
case of Abell~520 this finding also strengthens our confidence in the
detection of the ``dark'' peak 3.

There are a number of potential sources of systematics. One of these
results from errors in astrometry. First, scale variations can lead to
spurious shears. Secondly, when stacking data, the images need to be
aligned well in order to avoid anisotropies in the galaxy shapes.  For
our analysis we first ensure that the images are aligned well.  We
compare the RMS residuals in the positions of all non-saturated
pointlike sources within 15\arcmin of the X-ray center.  We find that
the Subaru $R$ and $i'$ images have an RMS deviation of $0.3$ pixels,
or $0.06\arcsec$ (0.2 kpc) with respect to the CFHT image. Differences
in the astrometric calibration are therefore negligible for the
purposes of our comparison of the lensing signal.

As described in \S\ref{sec:optdata} we measure the galaxy shapes from
two independent sets of Megacam $r'$ data. Similarly, we obtain shape
catalogs for each of the Subaru exposures (corrected for PSF
anisotropy). The next step is to create a master catalog per bandpass
by averaging the ellipticities of the galaxies that are found in
multiple catalogs. Note, however, that for large (and bright) galaxies
the error in the lensing signal is dominated by their intrinsic
shape. Therefore, combining their shape measurements improves the weak
lensing measurements only slightly. However, the results do improve
more for the faint galaxies, for which the measurements are dominated
by noise in the image.

The previous step results in three master catalogs, one for each
filter. We use this catalog to determine the aperture masses
\nocite{Hoekstra07}({Hoekstra} 2007) for each of the five peaks shown in
Figure~\ref{fig:bigpic}. However, the various selections (e.g.,
magnitude cuts, accuracy in the shape measurement) result in different
effective source redshift distributions. We used the photometric
redshift distributions derived by \nocite{Ilbert06}{Ilbert} {et~al.} (2006) to
estimate $\beta$ for each catalog. The resulting values are listed in
Table~\ref{tbl:beta}.

\begin{figure*}
\begin{center}
\resizebox{5in}{!}{\includegraphics{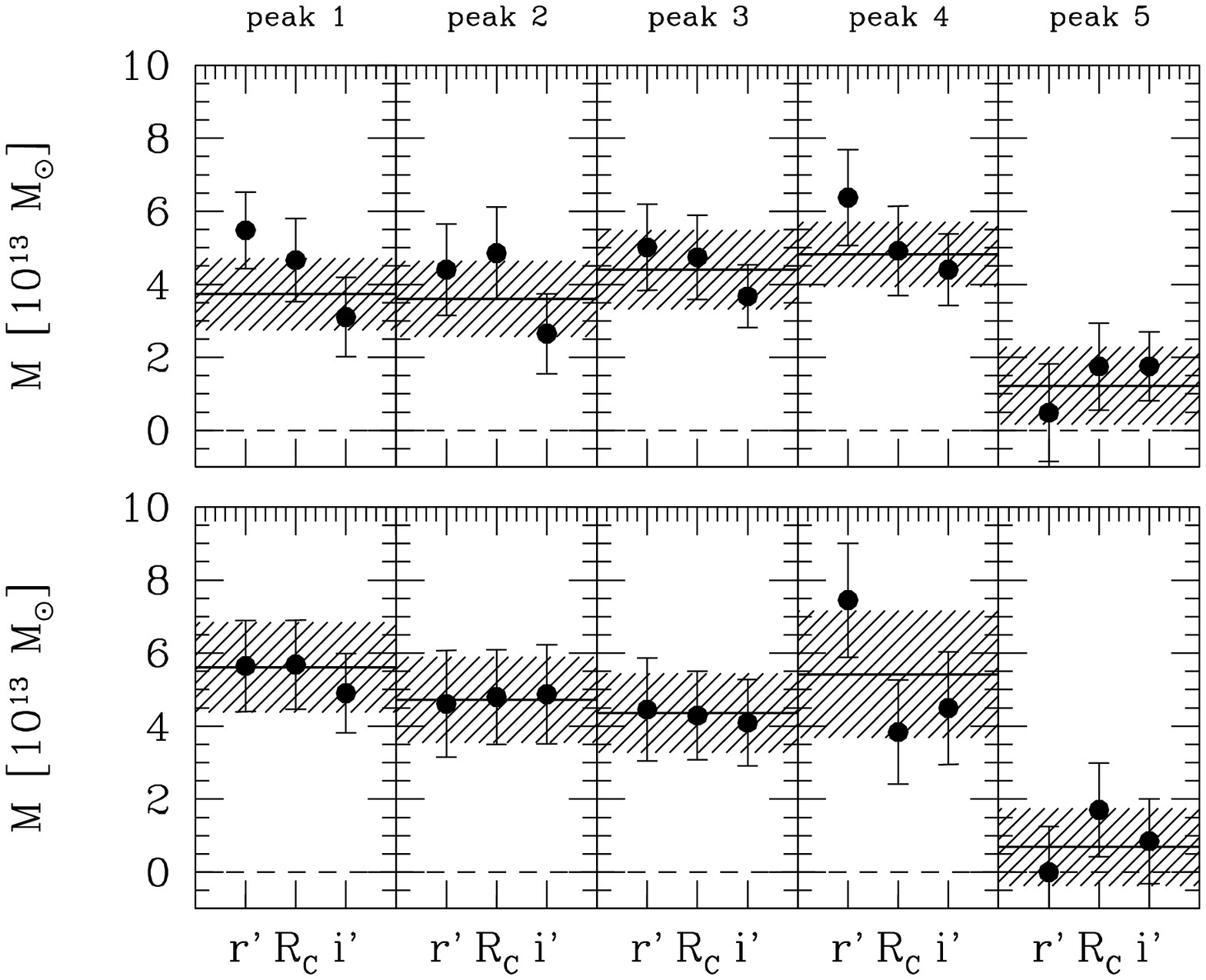}} \figcaption{Comparison
of aperture weak lensing masses for all detected galaxies (\emph{top
row}) and only galaxies detected in all three images (\emph{bottom
row}). The shaded region shows the 1$\sigma$ error bar for the
combined mass. The measured aperture masses are fully consistent
across all filters and both instruments.
\label{fig:mpeak}}
\end{center}
\end{figure*}

The points in the upper row of Figure~\ref{fig:mpeak} show the
resulting aperture masses (150 kpc radius) for each filter and
peak. The agreement between the estimates is excellent. Importantly, a
significant amount of matter is detected around peak 3, whereas little
signal is present at peak 5: we find the same result in both Subaru and
CFHT data, which have very different systematics. 

We also combine the three master catalogs into a single catalog which
is used to produce the mass reconstruction in Figure~\ref{fig:bigpic}
and mass estimates in Table~\ref{tbl:masses}. The masses obtained from
this catalog are indicated by the shaded areas in the top row of
Figure~\ref{fig:mpeak}.

Instead of considering all detected galaxies, we can limit the
analysis to galaxies detected in all three filters. In this case the
intrinsic shapes of the sources is almost completely removed from the
error budget.  The resulting masses are shown in the bottom panel of
Figure~\ref{fig:mpeak}. As before the shaded are is the average of the
three (matched) catalogs. As expected the variation from filter to
filter is very small, with somewhat larger variation for peak 4.  Note
that in this case the value for $\beta$ is almost filter independent
(see Table~\ref{tbl:beta}). 

Also in this case, a significant mass is inferred for peak 3, whereas
little mass is associated with peak 5.  Of the above methods, the use
of the unmatched catalog is the ``traditional'' approach, whereas the
matched catalogs provide a strong test of instrument/bandpass
cross-calibration: in the absence of instrumental bias, use of the
same exact galaxy catalog and technique ought to yield masses and mass
reconstructions with highly correlated signal and noise.

We find that the masses obtained using the matched catalogs are
somewhat higher compared to the unmatched catalogs. However, the
ratios are consistent with unity within the errors. The ratios of the
mass from the matched catalog to that of the unmatched sample for the
various filters and peaks are listed in Table~\ref{tbl:beta}. We omit
peak 5, because of its low mass, which leads to a meaningless ratio.
The difference in mass between the two catalogs may well be caused by
differences in the source redshift distributions which we did not
account for. Nevertheless, this study demonstrates that we can recover
the masses within $\sim 10\%$ using different filters and telescopes.

\begin{figure*}
\begin{center}
\resizebox{7in}{!}{\includegraphics{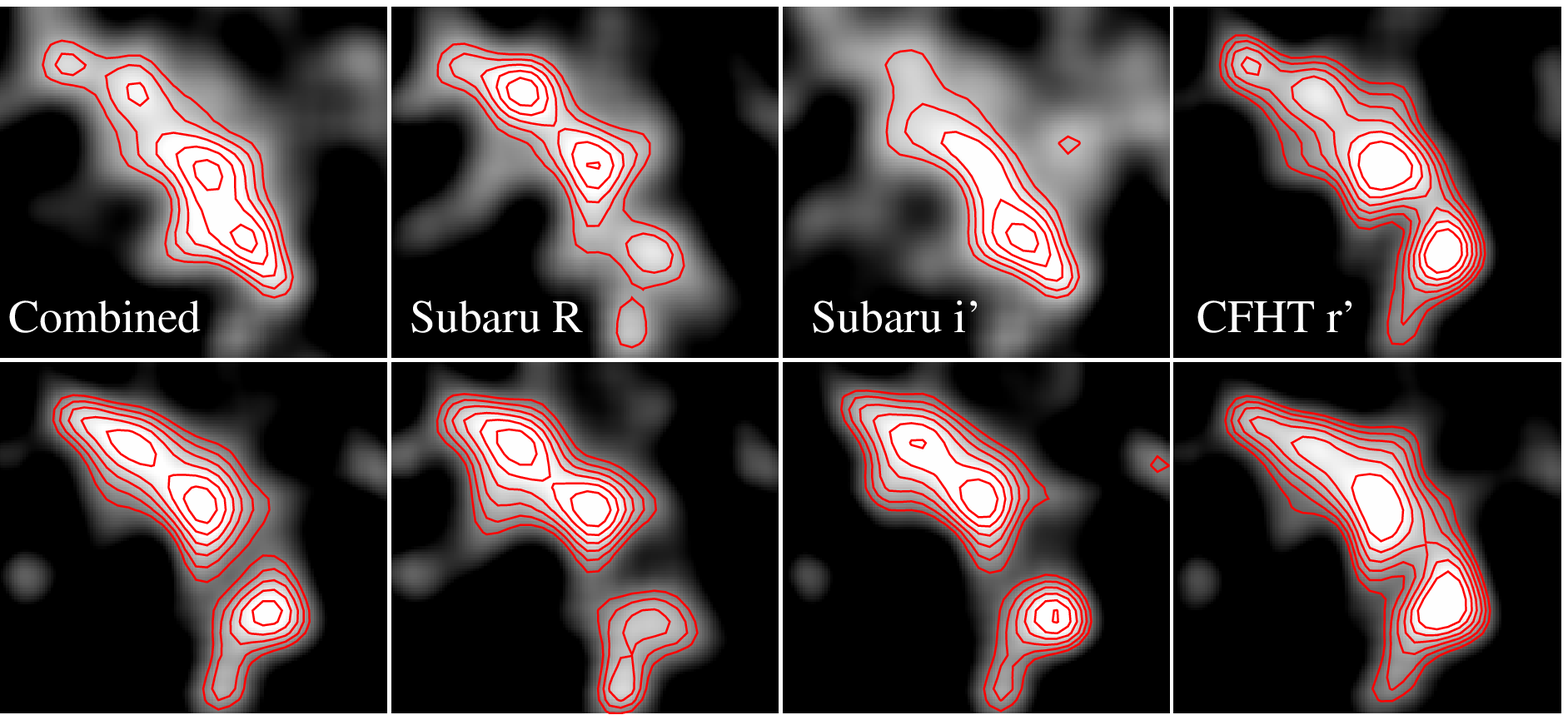}} \figcaption{Surface
mass reconstruction using all detected galaxies (\emph{top row}) and
using only galaxies detected in all three images (\emph{bottom
row}). Each column shows mass reconstructions using the Subaru $R$ and
$i'$ and the CFHT MegaCam $r'$ data; ``combined'' shows the mass
reconstruction using all three colors. The contours show convergence
levels equal to those in Figure 1a. We recover the central dark peak
in Abell 520 in all cases.
\label{fig:compare}}
\end{center}
\end{figure*}

Although the aperture mass estimates are of interest because of their
easy interpretation, it is also useful to compare the actual surface
mass density reconstructions.  The results of the individual mass
reconstructions are presented in Figure \ref{fig:compare}. The upper
row shows the results using the unmatched catalogs ($r'$, $R_c$, $i'$
and combined from left to right), whereas the lower panel show the
results for the matched catalogs.

The reconstructions are very similar, and the differences can be
attributed to noise. Of particular importance for our work is the
fact that we find that the dark peak (``peak 3'') is present in
all all datasets, regardless of whether we use only the matched
galaxies or not; similarly, none of the reconstructions shows an
overdensity at the position of peak 5.

These results differ from the \nocite{Okabe07}{Okabe} \& {Umetsu} (2007) reconstruction, who find
a significant mass at the position of peak 5. We believe that a
problem with the PSF anisotropy correction is a likely candidate,
given the fact that \nocite{Okabe07}{Okabe} \& {Umetsu} (2007) analyzed stacked data.  Based on
our consistent results for both the CFHT and Subaru data, we are
confident that the dark peak in Abell 520 is not the result of
instrumental effects, and that our methodology yields consistent
results in the $R$, $r'$, and $i'$ bandpasses.

\begin{deluxetable}{lccc|ccccc}
\trot
\tablewidth{0in}
\tablecaption{Object selection \label{tbl:beta}}
\tablehead{\colhead{filter} & \colhead{range} & $\beta$ & $\beta$ & & & mass ratio & \\ 
 & [mag] & (unmatched) & (matched) & cluster & peak 1 & peak 2 & peak 3 & peak 4}
\startdata
$r'$    & $20-25$ & 0.597 & 0.650 & $1.12\pm0.19$ & $1.03\pm0.30$ & $1.05\pm0.45$ & $0.89\pm0.35$ & $1.17\pm0.34$\\
$R_C$   & $20-25$ & 0.578 & 0.659 & $1.07\pm0.18$ & $1.22\pm0.39$ & $0.99\pm0.37$ & $0.90\pm0.34$ & $0.78\pm0.35$\\
$i'$    & $22-25$ & 0.648 & 0.660 & $1.02\pm0.15$ & $1.58\pm0.66$ & $1.84\pm0.92$ & $1.12\pm0.41$ & $1.02\pm0.42$\\
all     & $-$     & 0.589 & 0.653 & $1.11\pm0.17$ & $1.50\pm0.52$ & $1.31\pm0.50$ & $0.99\pm0.35$ & $1.12\pm0.41$\\
\enddata
\tablecomments{Shown are the magnitude selection for the sources in
the various filters, as well as the values for $\beta$ for each
selection. The table also lists the ratios of the inferred masses for
the matched and unmatched catalogs. The value of $\beta$ is computed
using the photometric redshift distribution from \nocite{Ilbert06}{Ilbert} {et~al.} (2006). The
variation of $\beta$ reflects the dependence on filter, but also
includes the down-weighting of faint galaxies, for which shapes cannot
be measured accurately.}
\end{deluxetable}

\ifthenelse{\equal{\mode}{aastex}}{ 
\section*{References}


}{
\bibliography{} }
\end{document}